\documentclass[aps,prc,reprint,amsmath,amssymb,showpacs,superscriptaddress]{revtex4-1}

\usepackage{CJK}
\usepackage{graphicx}
\usepackage{dcolumn}
\usepackage{bm}
\usepackage{color}
\usepackage{hyperref}
\usepackage{threeparttable}

\begin{document}
\begin{CJK*}{UTF8}{} 

\title{Principal components of nuclear mass models}

\author{X. H. Wu \CJKfamily{gbsn} (吴鑫辉)}
\email{wuxinhui@fzu.edu.cn}
\affiliation{Department of Physics, Fuzhou University, Fuzhou 350108, Fujian, China}
\affiliation{State Key Laboratory of Nuclear Physics and Technology, School of Physics, Peking University, Beijing 100871, China}

\author{P. W. Zhao \CJKfamily{gbsn} (赵鹏巍)}
\email{pwzhao@pku.edu.cn}
\affiliation{State Key Laboratory of Nuclear Physics and Technology, School of Physics, Peking University, Beijing 100871, China}

\begin{abstract}
The principal component analysis approach is employed to extract the principal components contained in nuclear mass models for the first time.
The effects coming from different nuclear mass models are reintegrated and reorganized in the extracted principal components.
These extracted principal components are recombined to build new mass models, which achieve better accuracy than the original theoretical mass models.
This indicates that the effects contained in different mass models can work together to improve the nuclear mass predictions with the help of the principal component analysis approach.
\end{abstract}

\maketitle

\end{CJK*}

\section{Introduction}

Nuclear masses are of fundamental importance in nuclear physics, as they can reflect many underlying physical effects of nuclear structure information~\cite{Lunney2003Rev.Mod.Phys.}.
Nuclear masses are also extremely important for astrophysics, as they are needed to extract the reaction energies that go into the calculations of all involved nuclear reaction rates in the stellar evolutions~\cite{Mumpower2016Prog.Part.Nucl.Phys., Li2019Sci.ChinaPhys.Mech.Astron., Jiang2021Astrophys.J., Wu2022Astrophys.J.152, Meng2011SCMPA}.
With the development of modern accelerator facilities, about 2500 nuclear masses have been measured so far~\cite{Wang2021Chin.Phys.C}.
Nevertheless, there is still a large uncharted territory in the nuclear landscape that cannot be accessed experimentally at least in the foreseeable future.

Theoretical prediction of nuclear mass has been a longstanding challenge of nuclear physics, due to the difficulties in understanding both the nuclear interactions and the quantum many-body systems.
Theoretical prediction of nuclear mass can be traced back to the macroscopic Weizs\"{a}cker mass formula based on the liquid drop model (LDM)~\cite{Weizsaecker1935Z.Physik}, which includes the bulk properties of nuclei quite well but lacks other effects.
Efforts have been made in pursuing extensions of the LDM to include more effects, which are known as the macroscopic-microscopic models~\cite{Wang2014Phys.Lett.B, Moeller2016Atom.DataNucl.DataTables, Koura2005Prog.Theor.Phys., Pearson1996Phys.Lett.B}.
The microscopic mass models have also been developed, based on the nonrelativistic and relativistic density functional~\cite{Geng2005Prog.Theor.Phys., Goriely2009Phys.Rev.Lett., Goriely2009Phys.Rev.Lett.a, Xia2018Atom.DataNucl.DataTables, Meng2020SCPMA, Erler2012Nature, Afanasjev2013Phys.Lett.B, Yang2021Phys.Rev.C, Zhang2022Atom.DataNucl.DataTables, Pan2022Phys.Rev.C, Sun2011SCMPA, Hua2012SCPMA, Qu2013SCPMA}.
In addition, the local mass relations such as the Garvey-Kelson relations~\cite{Barea2008Phys.Rev.C, Bao2017SCPMA}, the isobaric multiplet mass equation~\cite{Ormand1997Phys.Rev.C}, and the residual proton-neutron interactions~\cite{Fu2011Phys.Rev.C} are also used to predict masses of nuclei closed to the known region.

To precisely describe nuclear masses, one should in principle properly address all the underlying effects of nuclear quantum many-body systems, e.g., bulk effects, deformation effects, shell effects, odd-even effects, and even some unknown effects.
Different models may include these effects to different degrees.
Some models may properly consider several of these effects but improperly (less or over) consider several other effects, and some models may be otherwise.
With so many mass models at hand, one may ask: Can we extract the major patterns considered in these mass models? Can we refine the nuclear mass predictions by recombining the extracted patterns?

Recently, machine-learning approaches have attracted a lot of attention in physics and nuclear physics~\cite{Carleo2019Rev.Mod.Phys., Boehnlein2022Rev.Mod.Phys., He2023Sci.ChinaPhys.Mech.Astron., He2023NST, Ma2023CPL, Wang2023FoP, Alhassan2022NST, Zhou2024PPNP}, and have been successfully and widely employed in refining the predictions of nuclear masses, e.g., the kernel ridge regression (KRR)~\cite{Wu2020Phys.Rev.C051301, Wu2021Phys.Lett.B, Wu2022Phys.Lett.B137394, Guo2022Symmetry, Du2023Chin.Phys.C, Wu2023Front.Phys.}, the radial basis function (RBF)~\cite{Wang2011Phys.Rev.C, Niu2013Phys.Rev.C, Ma2017Phys.Rev.C, Niu2018Sci.Bull.}, the neural network (NN)~\cite{Utama2016Phys.Rev.C, Niu2018Phys.Lett.B, Neufcourt2018Phys.Rev.C, Niu2022Phys.Rev.C, Ming2022NST}, the Gaussian process regression~\cite{Neufcourt2019Phys.Rev.Lett., Shelley2021Universe}, the Levenberg-Marquardt neural network~\cite{Zhang2017J.Phys.GNucl.Part.Phys.}, the light gradient boosting machine~\cite{Gao2021Nucl.Sci.Tech.}, the Bayesian probability classifier~\cite{Liu2021Phys.Rev.C},
the probably approximately correct learning~\cite{Idini2020Phys.Rev.Research},
etc.
These successes encourage us to employ statistical techniques to analyze the major effects contained in nuclear mass models.
Principal component analysis (PCA) is a popular statistical technique for feature selection and dimensionality reduction~\cite{Wold1987PCA, Jolliffe2002principal}.
PCA has been applied to the studies of nuclear physics, e.g.,
to remove spurious events in the measurement of the 2$\nu$$\beta\beta$ decay~\cite{Augier2023PRL},
to evaluate the uncertainty in the neutrino-nucleus scattering cross section~\cite{Akimov2022PRL},
to study event-by-event fluctuations in relativistic heavy-ion collisions~\cite{Bhalerao2015PRL},
to reduce dimensions of many-body problem~\cite{Bonilla2022PRC},
to optimize functional derivative of nuclear energy density functional~\cite{Wu2022Phys.Rev.C},
to analyze the correlation of parameters in nuclear energy density functional~\cite{Bulgac2018Phys.Rev.C},
to quantify uncertainty of empirical shell-model interaction~\cite{Fox2020PRC},
to learn about the number of effective parameters of the liquid drop model and the Skyrme functional~\cite{Kejzlar2020JPG},
to define empirical basis functions capturing the variation in the output of Hartree-Fock-Bogoliubov calculations~\cite{Schunck2020JPG},
etc.

In this work, for the first time, the PCA approach is employed to extract the principal components contained in several widely used nuclear mass models.
The commonalities and differences across different mass models are analyzed with the help of these principal components.
These principal components are then recombined to build new mass models.
This is different from the existing work of applying PCA to nuclear mass study~\cite{Kejzlar2020JPG}, where the PCA is employed to learn about the number of effective parameters of the liquid drop model and the Skyrme functional.

\section{Theoretical framework}

The PCA is a technique to identify a set of principal components that capture the maximum features in the data~\cite{Wold1987PCA, Jolliffe2002principal}.
This is achieved by transforming the origin variables to a new set of variables, the principal components, which are uncorrelated, and which are ordered so that the first few retain most of the features present in all of the original variables.

When applying the PCA to nuclear mass models, the original variables are the mass predictions of different nuclear models, i.e., original nuclear mass tables.
These original variables are correlated.
They share lots of effects in common.
After performing the PCA, they are transformed into a new set of ``principal mass models'', which are uncorrelated and arranged in the order of importance in representing the relevant features extracted from the original mass models.

The PCA application to nuclear mass models includes the following steps.
\begin{itemize}
  \item First, pick up $N$ mass models that would be analyzed, e.g., model-1, model-2, $\cdots$, model-$N$.
  \item Second, vectorize mass predictions of these mass models into high-dimension vectors with $m$ components (corresponding to $m$ nuclei in the nuclear chart), i.e., $\bm{M}_1$, $\bm{M}_2$, $\cdots$, $\bm{M}_N$, namely ``original mass-model vector".
  \item Third, construct the covariance matrix of these $N$ original mass-model vectors, $\bm{C}=\bm{X}^{\top}\bm{X}$, where, $\bm{X}$ is a $N\times m$ matrix defined as $\bm{X} = (\bm{M}_1, \bm{M}_2, \cdots, \bm{M}_N)^{\top}$. Since the covariance matrix is constructed from $N$ original mass-model vectors, the rank of the covariance matrix is $N$.
  \item Fourth, diagonalize the covariance matrix $\bm{C}$, which gives $N$ non-trivial eigenvectors $\bm{v}_i$, namely ``principal mass-model vector", that are listed in decreasing order with the eigenvalues $\lambda_i$.
\end{itemize}

Following these standard steps of the PCA, one obtains the so-called principal components (PCs), represented by the principal mass-model vectors, i.e., $\bm{v}_1, \bm{v}_2, \cdots, \bm{v}_N$.
Each dimension of a mass-model vector corresponds to a nucleus in the nuclear chart.
The importance of a mass-model vector is represented by the corresponding eigenvalues $\lambda_i$.
This means that $\bm{v}_1$ represents the most important relevant feature extracted from these $N$ mass models, $\bm{v}_2$ is the second important one, and so on, while $\bm{v}_N$ is the last relevant one.
Each principal component is a pattern of mass predictions for nuclei over the nuclear chart.
With these PCs of nuclear mass models, one can better analyze the commonalities and differences across different mass models and build new mass models.

\section{Numerical details}

Six mass models, i.e., FRDM2012~\cite{Moeller2016Atom.DataNucl.DataTables}, HFB17~\cite{Goriely2009Phys.Rev.Lett.}, KTUY05~\cite{Koura2005Prog.Theor.Phys.}, D1M~\cite{Goriely2009Phys.Rev.Lett.a}, RMF~\cite{Geng2005Prog.Theor.Phys.}, LDM~\cite{Weizsaecker1935Z.Physik}, are picked up to extract principal components.
The overlap of these mass models includes 6254 nuclei, therefore each mass-model vector, as well as each principal component, are vectors in 6254-dimension Hilbert space.
The mass data from AME2020~\cite{Wang2021Chin.Phys.C} are adopted to evaluate the principal components.
The overlap of these mass models with the experimental data AME2020~\cite{Wang2021Chin.Phys.C} includes 2421 nuclei, which constitutes a subspace of the 6254-dimension Hilbert space.
Since six mass models are considered, the components with $\lambda_i$ smaller than $\lambda_6$ are irrelevant.
Therefore, one obtains six principal components (PCs) extracted from these mass models, i.e., $\bm{v}_1$ for principal component-1 (PC1), $\bm{v}_2$ for PC2, and so on.

\section{Results and discussion}

\begin{figure}[!htbp] 
\centering
\includegraphics[width=8.5cm]{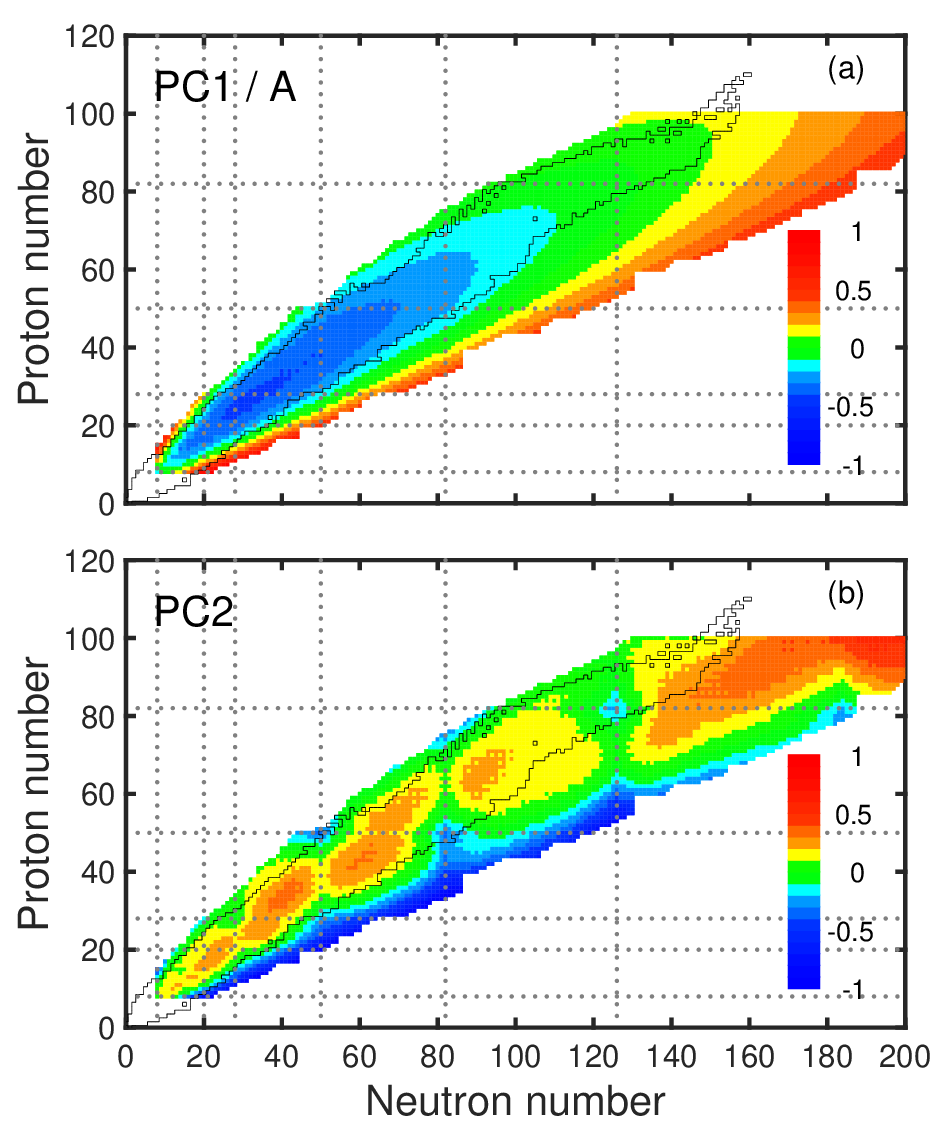}
\caption{The principal components, i.e., PC1 (a) and PC2 (b), of nuclear mass models with the values scaled to the range between $-1$ and $1$.
The PC1 is scaled with the mass number $A$ to better illustrate the most bound nuclei around the iron-group elements.
The boundary of nuclei with known masses in AME2020 is shown by the black contour lines.
Dotted lines indicate the magic numbers.
The inset of panel (b) presents PC2 of the Nd isotope chain.
}
\label{fig1}
\end{figure}

\begin{figure}[!htbp]
\centering
\includegraphics[width=8.5cm]{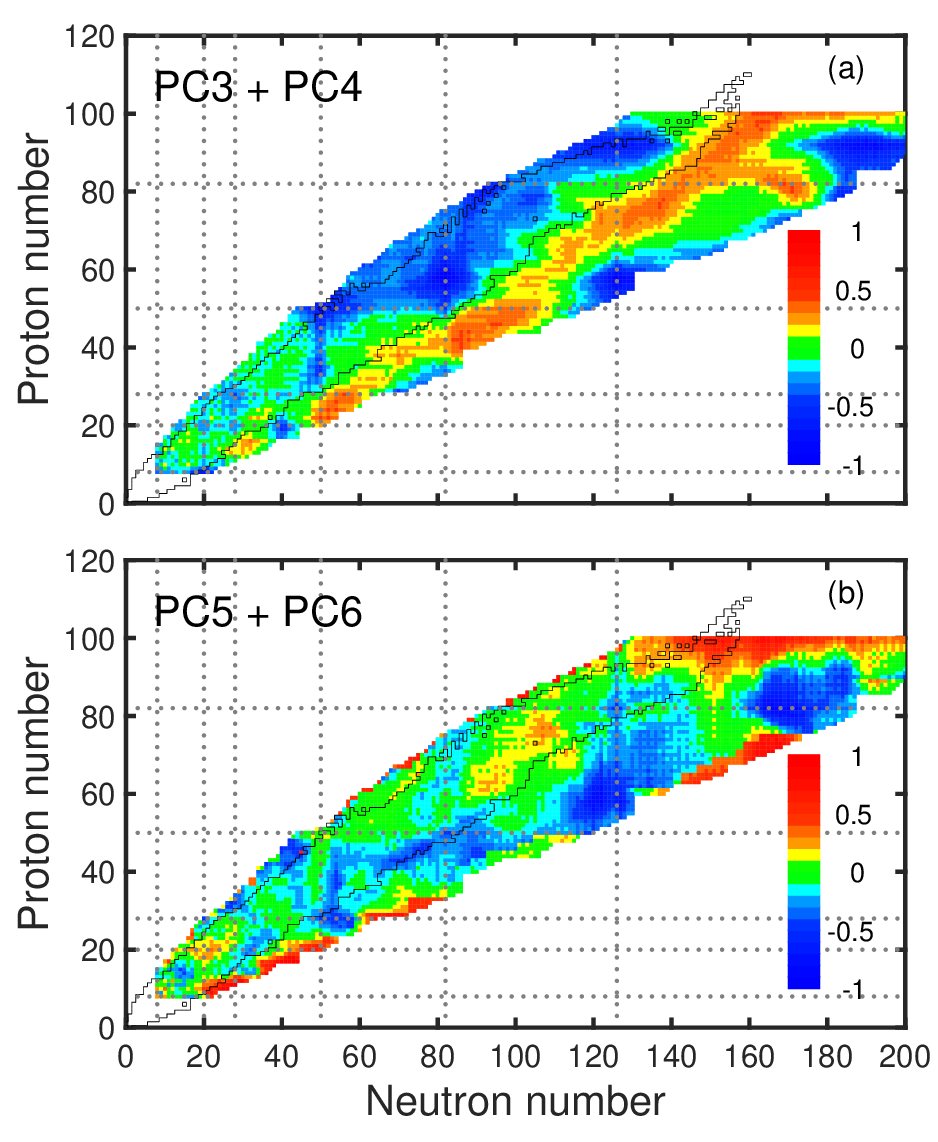}
\caption{Similar with Fig.~\ref{fig1}, but for PC3, PC4, PC5, and PC6.
The PC3 and PC4 are added together with the weights determined by the eigenvalues, same for PC5 and PC6.}
\label{fig2}
\end{figure}

The eigenvalues corresponding to the six principal components are presented in the second row of Table~\ref{tab1}, together with overlaps of the six mass models (original mass-model vectors) with six principal components (principal mass-model vectors) which are presented from 3rd to 8th rows.
The visualizations of the six PCs are shown in Fig.~\ref{fig1} and Fig.~\ref{fig2}.
Each principal component is a pattern of mass predictions for nuclei over the nuclear chart.
Since each principal component is an eigenvector of the covariance matrix, the scale value of the principal component is free.
For convenience, the values presented in Fig.~\ref{fig1} and Fig.~\ref{fig2} are scaled to the range between $-1$ and $1$.

\begin{table*}[!t]
\caption{The corresponding eigenvalues of the six principal components extracted from the six mass models, and the overlaps* of the six principal components with the six mass models and the experimental data AME2020.}
\begin{center}
\setlength{\tabcolsep}{3.5mm}{
\begin{tabular}{c c c c c c c}
\hline
\hline
 & PC1 & PC2 & PC3 & PC4 & PC5 & PC6 \\
\hline
 Eigenvalues & $6.2\times10^{10}$ & $3.5\times10^7$ & $7.7\times10^6$ & $4.3\times10^6$ & $1.0\times10^6$ & $6.4\times10^5$ \\
\hline
 FRDM2012    & 0.99985 & 0.01275 & -0.00773 & -0.00267 & -0.00781 & -0.00087 \\
\hline
 HFB17       & 0.99983 & 0.01497 & 0.00038 & -0.00768 & 0.00421 & -0.00056 \\
\hline
 KTUY05      & 0.99986 & 0.01228 & 0.00099 & -0.00890 & 0.00231 & 0.00613 \\
\hline
 D1M         & 0.99980 & 0.00023 & -0.01488 & 0.01274 & 0.00291 & 0.00062 \\
\hline
 RMF         & 0.99968 & 0.00926 & 0.02160 & 0.00955 & -0.00101 & -0.00003 \\
\hline
 LDM         & 0.99851 & -0.05428 & 0.00204 & -0.00347 & -0.00043 & -0.00040 \\
\hline
 AME2020**  &  0.99987 & 0.01053 & -0.00760 & -0.00264 & -0.00032 & -0.00164  \\
\hline
\hline
\end{tabular}}
\begin{tablenotes}
* The overlap of the principal component $\bm{v}_i$ with the original mass model $\bm{M}_j$ is defined by $\frac{\bm{M}_j\cdot\bm{v}_i }{\sqrt{|\bm{M}_j ||\bm{v}_i |}} = \frac{\sum_{k}^{m} M_j(k) v_i(k)}{\sqrt{\sum_{k}^{m} M^2_j(k)}\sqrt{\sum_{k}^{m} v^2_i(k)}} $. \\
** Note that when the Hilbert space is reduced from 6254 to 2421 dimensions, the six PCs are not orthogonal to each other anymore.
Therefore, they are re-orthogonalized with the Schmidt orthogonalization before calculating the overlaps with the AME2020.
\end{tablenotes}
\end{center}
\label{tab1}
\end{table*}

By comparing the eigenvalues of different PCs in Table~\ref{tab1}, it can be seen that the eigenvalue $6.2\times10^{10}$ corresponding to the PC1 is much larger than other ones, which means it is the most important component of the mass models that contribute to the major part of nuclear masses.
As is known, the bulk properties contribute to the major part of nuclear masses.
The bulk properties were originally described by the LDM model, including volume term, surface term, Coulomb term, symmetry energy term, and odd-even term.
Of course, these properties are also managed well in other mass models.
Therefore, the major contributions of the PC1 extracted from different mass models should correspond to the bulk properties.
This can be seen clearly in Fig.~\ref{fig1}~(a), where the most bound nuclei located around the iron-group elements as been described by all nuclear mass models since the LDM.
The large eigenvalue of the PC1 also indicates the large similarity of different mass models.
This can be seen from the overlaps of the six mass models with PC1 (second column of Table~\ref{tab1}), which are similar and around 0.999.

Things are different when looking at the other PCs.
As can be seen in Table~\ref{tab1}, their eigenvalues are much smaller than the one of PC1, and their overlaps with different mass models are relatively small and no longer similar to each other.
The eigenvalue $3.5\times10^7$ corresponding to PC2 is obviously larger than the latter PCs, which represents the second important pattern contained in the nuclear mass models.
The visualization of PC2 is illustrated in Fig.~\ref{fig1}~(b).
One prominent feature of the PC2 is the deformation properties related to the shell effects.
The related magic numbers predicted by the PC2 are the same as the traditional ones, which can be seen clearly in Fig. 1 (b) with the help of the magic lines.
Another feature that can be seen in Fig.~\ref{fig1}~(b) is the grain structure, which can be seen more clear in the inset of Fig.~\ref{fig1}~(b) as the odd-even staggering behaviors. This corresponds to the odd-even effects originally from the pairing correlations and Pauli blocking.
One can find that the overlap of the LDM model with the PC2 is opposite to the overlaps of the other five models.
This is because the deformation effects are included in all other five mass models except for the LDM model.

The eigenvalues of PC3 and PC4 are small and close to each other, therefore these two principal components are added together in Fig.~\ref{fig2}~(a).
The significant structure of PC3+PC4 is the different behavior between the neutron-rich and the proton-rich sides, i.e., more bound for one side and less bound for the other side.
This might be understood as the correction of the breaking of neutron and proton symmetry energy.
Several blocky structures divided by the magic lines and grain structures can be also seen in Fig.~\ref{fig2}~(a), which indicates that some residual deformation and odd-even effects are also contained in the PC3+PC4.
The eigenvalues of PC5 and PC6 are even smaller, so their contributions are smaller compared to the first four principal components.
The patterns of PC5 and PC6 as shown in Fig.~\ref{fig2}~(b) are much more irregular.
As would be seen later, these two components only contribute about 10-keV improvement to reproduce the mass data AME2020.

\begin{figure}[!htbp]
\centering
\includegraphics[width=8.5cm]{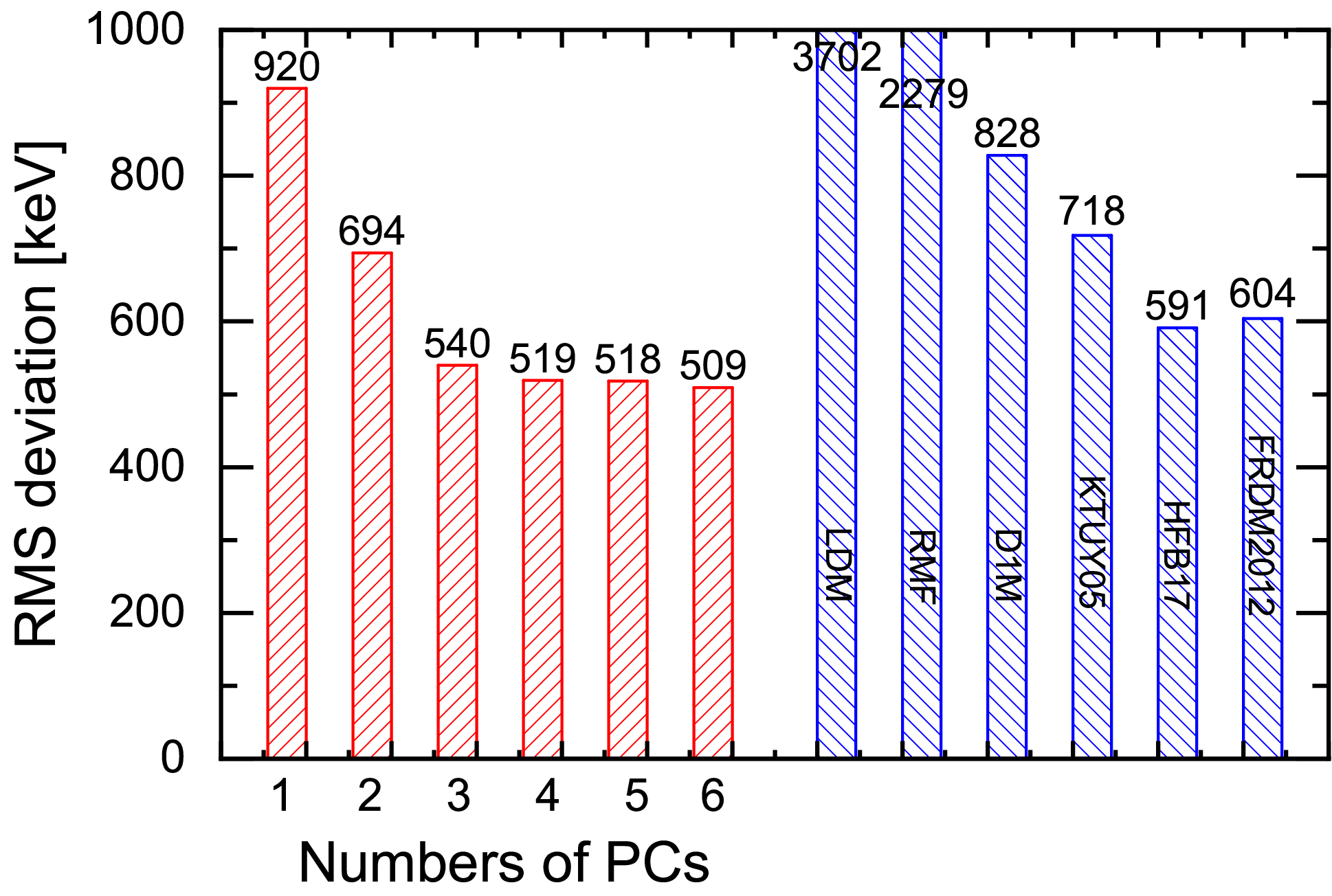}
\caption{Root-mean-square (rms) deviations between new mass models including different numbers of PCs and the experimental data AME2020.
The rms deviations between the six origin mass models and the experimental data AME2020 are also presented for comparison.}
\label{fig3}
\end{figure}

With these extracted principal components of nuclear mass models at hand, one can recombine these PCs to build new mass models.
The superposition coefficients are determined by the overlaps of these PCs with the mass data, which are presented in Table~\ref{tab1}.
The overlaps represent the contributions of these components to reproduce the data, and they generally decrease with the decreasing of eigenvalues of these PCs.
Note that the PCs and their sort orders are extracted from the theoretical mass models, without any information from the experimental data AME2020.
Therefore, it is appealing to find that the order of contributions to reproduce the experimental data is the same as the order of the corresponding eigenvalues of these components.
This means that the major effects of nuclear masses have been well captured with the efforts of different mass models.
With these PCs and the corresponding superposition coefficients, new mass models can be built by including several of these principal components, i.e., from one PC to six PCs.

The root-mean-square (rms) deviations between new mass models including different numbers of PCs and the experimental data AME2020 are presented in Fig.~\ref{fig3}.
As can be seen, the more PCs being included the more precise can be achieved.
The inclusion of the first principal component can already achieve a precision of 920 keV.
The inclusions of the second, third, and fourth components work well to further improve the precision of reproducing the data.
It should be noted that, with the inclusions of the first four principal components, the new model achieves a precision of 519 keV, which is already smaller than the smallest one of the six original theoretical mass models, i.e., 591 keV of the HFB17 model.
If all six PCs are included, the rms deviation can be reduced to 509 keV.
This means that, by recombining the principal components extracted from the theoretical mass models, one can build a better mass model than the original theoretical models.
It should be reminded that the information or effects contained in the principal components are also contained in the six mass models.
However, these effects are reintegrated by the PCA technology, which means that the effects coming from different mass models are reorganized to build new nuclear mass models.
Therefore, the results indicate that the effects included in different theoretical mass models can work together to improve the prediction of nuclear masses, and this can be done by the PCA technology.
Note that based on the six original mass models, one can also construct new mass models by arithmetic averaging or weighted averaging (with weights being inversely proportional to the rms deviations with respect to the experimental data) these six models. The corresponding rms deviations obtained by these two averaging mass models are 913 and 525 keV respectively, which are both larger than 509 keV.

To further examine this conclusion, the RCHB~\cite{Xia2018Atom.DataNucl.DataTables} and WS4~\cite{Wang2014Phys.Lett.B} models are added to the original six mass models respectively, and the PCA is performed for these two sets of seven mass models.
The RCHB model is a nuclear mass model based on the relativistic density functional theory with continuum effects but limited to the assumption of spherical symmetry~\cite{Xia2018Atom.DataNucl.DataTables}.
The rms deviation of the RCHB mass model with the experimental data is 7960 keV~\cite{Xia2018Atom.DataNucl.DataTables}, which is larger than all the ones of the six mass models.
When the RCHB model is included, the rms deviation of the new mass model constructed by seven PCs with the experimental mass data is 456 keV, which is obviously smaller than 509 keV.
This indicates that, although the rms deviation of the RCHB model is large due to the spherical symmetry assumption, the effects included in the RCHB mass model can still help to improve the predictions of nuclear masses.
These effects can be extracted by the PCA technology.
The rms deviation of the WS4 mass model with the experimental data is as small as 298 keV~\cite{Wang2014Phys.Lett.B}.
When the WS4 model is included, the rms deviation of the new mass model including seven PCs can be reduced to 292 keV, significantly smaller than 509 keV and slightly smaller than 298 keV.
This means that the effects included in the WS4 mass model can significantly help the six models to build a better mass model, and the effects in the six models can also help the WS4 model.

To examine the reliability of the mass model constructed by the PCs in predicting unknown regions, an extrapolation validation is performed. In the extrapolation validation, the superposition coefficients of the PCs are determined by experimental data with only $Z\leq60$ nuclei being included. The new mass model constructed by the PCs with these coefficients achieves a precision of 549 keV in describing all the experimental data, which is still at the same lever with 509 keV. This indicates that the PCA works well to avoid the overfitting problem. This is because the new mass model is constructed by the principal components extracted from theoretical mass models and its reliability is guaranteed by the major effects included in these theoretical models. It is thus natural to believe the reliability of the new mass model in the experimentally unknown region.

Note that the Bayesian model averaging (BMA) method~\cite{Neufcourt2019Phys.Rev.Lett., Zhang2024NPA} is somehow similar to the PCA method introduced in this work. Both of them construct new nuclear mass models in the representation space of selected nuclear mass models and aim to combine the advantages of different nuclear models. However, their differences are also very clear. For the BMA method, the new mass model is constructed by weighted averaging of the selected models, where the weights are determined based on Bayes theorem.
The BMA has the advantage that it can help identify better models and discard poor models with the determined weights. For the PCA method, the new mass model is constructed by superpositions of principal components of nuclear mass models. These principal components are extracted from the theoretical mass models, and they represent the major effects included in the selected mass models. The advantage of PCA is that it can extract the major effects included in the theoretical mass models and arrange them in the order of importance. Therefore, one can know what are the most important effects contained in nuclear mass models and build a new precise mass model with only several principal components.

\section{Summary}

In summary, the principal component analysis approach is employed to extract the principal components contained in nuclear mass models.
The effects coming from different nuclear mass models are reintegrated and reorganized in the extracted principal components by the PCA technology.
The first principal component of nuclear masses is mainly contributed by the bulk property as described in the LDM, and the second principal component is mainly contributed by the deformation related to shell effects and the odd-even effects
The breaking of neutron and proton symmetry energy is also an important component that contributes to the nuclear masses, which are included in the third and fourth principal components.
These extracted principal components are then recombined to build new nuclear mass models.
It is found that new mass models can achieve better accuracy in predicting the experimental mass data than the original theoretical mass models.
This indicates that the effects contained in different theoretical models can work together to improve the nuclear mass predictions, and this can be done by the PCA technology.

This can provide a new manner to build nuclear mass models by extracting the principal components of different nuclear mass models and then recombining these components.
The fully \textit{ab initio} calculations that can in principle include all effects, which are based on exact nuclear interaction and many-body calculation without approximation, are extremely difficult (if not impossible) for nuclear masses all over the nuclear chart even in the foreseeable future.
Therefore, it can be said that all nuclear mass models, including the ones that already exist and the ones that might be built in the future, contain some kinds of approximations or ignore some kinds of effects.
In other words, they include different kinds and different degrees of effects of the nuclear masses.
It therefore would be interesting to try another way rather than fully \textit{ab initio} calculations to build nuclear mass tables including as many effects as possible by extracting and combining the effects included in different models.
This work shows that the PCA technology can work as one candidate for this purpose.
We also encourage theorists to develop new theoretical mass models with new effects and do not need to worry too much about the balance of including new effects and the accuracy of reproducing experimental data, because the PCA technology can dig out the new effects and make them contribute to improve the nuclear mass predictions.


\begin{acknowledgments}
This work was partly supported by the State Key Laboratory of Nuclear Physics and Technology, Peking University (Grant No. NPT2023KFY02), the China Postdoctoral Science Foundation (Grant No. 2021M700256), the National Key R\&D Program of China (Contract No. 2018YFA0404400), the National Natural Science Foundation of China (Grants No. 11935003, No. 11975031, No. 12141501, No. 12070131001), and the High-performance Computing Platform of Peking University.
\end{acknowledgments}

\end{document}